\documentclass[showpacs,aps,pre,superscriptaddress,eqsecnum]{revtex4}
\usepackage{graphicx}
\usepackage{amsfonts}
\usepackage{amssymb}
\usepackage{amsmath}
\usepackage{color, verbatim}

\newcommand{\sech}{\rm sech}
\newcommand{\ds}{\displaystyle}
\newcommand{\cnrk}{{\rm cn}(\rho,\kappa)}
\newcommand{\snrk}{{\rm sn}(\rho,\kappa)}
\newcommand{\dnrk}{{\rm dn}(\rho,\kappa)}
\newcommand{\cnSrk}{{\rm cn}^2(\rho,\kappa)}
\newcommand{\snSrk}{{\rm sn}^2(\rho,\kappa)}
\newcommand{\dnSrk}{{\rm dn}^2(\rho,\kappa)}
\newcommand{\cnxk}{{\rm cn}(x,\kappa)}
\newcommand{\snxk}{{\rm sn}(x,\kappa)}
\newcommand{\dnxk}{{\rm dn}(x,\kappa)}
\newcommand{\cnSxk}{{\rm cn}^2(x,\kappa)}
\newcommand{\snSxk}{{\rm sn}^2(x,\kappa)}
\newcommand{\dnSxk}{{\rm dn}^2(x,\kappa)}

\newcommand{\dnaxk}{{\rm dn}(a(x-vt),\kappa)}
\newcommand{\cnSaxk}{{\rm cn}^2(a(x-vt),\kappa)}
\newcommand{\snSaxk}{{\rm sn}^2(a(x-vt),\kappa)}
\newcommand{\dnSaxk}{{\rm dn}^2(a(x-vt),\kappa)}
\newcommand{\cn}{{\rm cn}}
\newcommand{\sn}{{\rm sn}}
\newcommand{\dn}{{\rm dn}}
\newcommand{\rf}[1]{(\ref{#1})}
\newcommand{\SR}{\stackrel}
\newcommand{\beq}{\begin{equation}}
\newcommand{\eeq}{\end{equation}}
\newcommand{\beqa}{\begin{eqnarray}}
\newcommand{\eeqa}{\end{eqnarray}}

\begin{document}

\title[Elliptic function solutions in Jackiw-Teitelboim dilaton gravity]{Elliptic function solutions in Jackiw-Teitelboim dilaton gravity}

\author{Jennie D'Ambroise}
\affiliation{ Department of Mathematics/CIS, SUNY Old Westbury, PO Box 210, Old Westbury, NY, 11568}
\email{dambroisej@oldwestbury.edu}

\author{Floyd L. Williams}
\affiliation{ Department of Mathematics and Statistics, University of Massachusetts, Amherst, MA, 01003}
\email{williams@math.umass.edu}

\begin{abstract}
We present a new family of solutions for the Jackiw-Teitelboim model of two-dimensional gravity with a negative cosmological constant.  Here, a metric of constant Ricci scalar curvature is constructed, and explicit linearly independent solutions of the corresponding dilaton field equations are determined.  The metric is transformed to a black hole metric, and the dilaton solutions are expressed in terms of Jacobi elliptic functions.  Using these solutions we compute, for example, Killing vectors for the metric.
\end{abstract}


\maketitle

\section{Introduction}

It is well-known that the Einstein  gravitational field equations for a vacuum (with a zero matter tensor) are automatically solved by any metric $g$ on a two-dimensional space-time $M$.  A proof of this fact is given in section 2 of \cite{1}, for example.  A non-trivial theory of gravity for such an $M$ was worked out in 1984 by R. Jackiw and C. Teitelboim (J-T).  This involves in addition to $g$ a scalar field $\Phi$ on $M$ called a dilaton field; see \cite{2},\cite{3}.  The pair $(g,\Phi)$ is subject to the equations of motion
\beq
R(g) = \frac{2}{l^2}, \qquad \nabla_i \nabla_j\Phi = \frac{g_{ij}\Phi}{l^2}\label{1.1}
\eeq
derived from the action integral 
\beq
S(g,\Phi) = \mbox{constant}\cdot \ds\int_M d^2x \sqrt{|\mbox{det} g|}\Phi \left(R(g) - \frac{2}{l^2}\right)\label{1.2}
\eeq
where $R(g)$ is the constant Ricci scalar curvature of $g$, and where the (negative) cosmological constant is $\Lambda = -1/l^2$.  In local coordinates $(x_1,x_2)$ on $M$, the Hessian in \rf{1.1} is given by 
\beq
\nabla_i\nabla_j\Phi = \frac{\partial^2 \Phi}{\partial x_i \partial x_j} - \ds\sum_{k=1}^2 \Gamma_{ij}^k \frac{\partial \Phi}{\partial x_k}, \quad 1\leq i,j \leq 2,\label{1.3}
\eeq
where $\Gamma_{ij}^k$ are the Christoffel symbols (of the second kind) of $g$ \cite{1}.  The J-T theory has, for example, the (Lorentzian) black hole solution
\beq
g: \ \ ds^2 = -(m^2r^2-M)dT^2 + \frac{dr^2}{m^2r^2-M}, \label{1.4}
\eeq
with coordinates $(x_1,x_2) = (T,r)$, where
\beq
\Lambda = -m^2,\quad R(g) = 2m^2, \quad \Phi(T,r) \SR{def.}{=}mr,\label{1.5}
\eeq
with $M$ a black hole mass parameter.  Here and throughout we note that our sign convention for scalar curvature is the negative of that used in \cite{2},\cite{3}, and by others in the literature.

The purpose of this paper is the following.  For real numbers $a,b\neq 0$ and for a soliton velocity parameter $v$ we consider the following metric in the variables $(x_1,x_2) = (\tau, \rho)$:
\beqa
 ds^2  \SR{def.}{=} a^2 b^2 \dnSrk \left[ \left( \frac{a^2 \kappa^4 \snSrk \cnSrk }{\dnSrk } - \frac{v^2}{4}\right)d\tau^2\right. \nonumber \\
 \left. - \frac{ \kappa^4 \snSrk \cnSrk }{\dnSrk } \left( \frac{a^2 \kappa^4 \snSrk \cnSrk }{\dnSrk } - \frac{v^2}{4}\right)^{-1} d\rho^2 
\right]\label{1.6}\eeqa
where $\snxk, \cnxk , \dnxk $ are the standard Jacobi elliptic functions with modulus $\kappa$; $0\leq \kappa \leq 1$ \cite{4}.  We will generally assume that 
\beq
\kappa \neq 0, \qquad \left| \frac{v}{2a\kappa^2}\right| > 1.\label{1.7}
\eeq
As will be seen later, this metric is the diagonalization of a metric constructed from solutions $r(x,t), s(x,t)$ of the \emph{reaction diffusion system}
\beqa
r_t - r_{xx} + \frac{2}{b^2}r^2s &=& 0\label{1.8}\\
s_t + s_{xx} - \frac{2}{b^2}rs^2 &=& 0.\nonumber
\eeqa
We will explicate the solutions $r(x,t), s(x,t)$ in terms of the elliptic function $\dnxk $.  Remarkably, the metric in \rf{1.6} has constant scalar curvature $R(g) = 8/b^2$ so that the first equation in \rf{1.1} holds.  The main work of the paper then is to solve the corresponding system of partial differential equations (the dilaton field equations) in \rf{1.1}, which for $g$ in \rf{1.6} are
\beq
\nabla_i \nabla_j \Phi = \frac{R(g)}{2} g_{ij}\Phi = \frac{4}{b^2} g_{ij}\Phi, \quad 1\leq i,j \leq 2.  \label{1.9}
\eeq
Here the cosmological constant is $\Lambda = -{4}/{b^2}$.

Given the complicated nature of our $g$, the system \rf{1.9} is necessarily quite difficult to solve directly.  Our method is to construct a series of transformations of variables so that $g$ in \rf{1.6} is transformed to $g$ in \rf{1.4}.   Then we can use the simple solution $\Phi(T,r) = mr$ in \rf{1.5}, and other known solutions, to work backwards through these transformations of variables to construct $\Phi(\tau,\rho)$ that satisfies \rf{1.9}.  The various details involved, with further remarks that lead to \rf{1.6}, will be the business of sections 2, 3, and 4.

In the end, we obtain the following main result:  The metric in \rf{1.6} solves the first J-T equation of motion \rf{1.1}.  Namely $R(g) = 8/b^2$, as we have remarked.  Also three linearly independent solutions of the field equations in \rf{1.1}, namely of the system of equations \rf{1.9}, are given by
\beqa
\Phi^{(1)}(\tau,\rho) &=& 2a^2 \dnSrk  + \frac{v^2}{4} - a^2(2-\kappa^2)\label{1.10}\\
\Phi^{(2)}(\tau,\rho) &=& \dnrk\sinh\left(\sqrt{A}\tau\right)\sqrt{\frac{v^2}{4} - a^2\kappa^4 \snSrk\cnSrk/\dnSrk}\nonumber\\
\Phi^{(3)}(\tau,\rho) &=& \dnrk\cosh\left(\sqrt{A}\tau\right)\sqrt{\frac{v^2}{4} - a^2\kappa^4 \snSrk\cnSrk/\dnSrk}\nonumber
\eeqa
for
\beqa
A&\SR{def.}{=}& \frac{v^4}{16} - \frac{v^2a^2}{2}(2-\kappa^2) + a^4\kappa^4,\label{1.11}
\eeqa
which we assume is non-zero.  Given \rf{1.7} we shall see in section 4 that $A=0$ only for $a=\pm(1-\sqrt{1-\kappa^2})v/2\kappa^2$ and moreover that the second expression under the radical (i.e. $v^2/4 - \cdots$) in \rf{1.10} is positive.  For $\kappa=1$, $A=(v^2/4 - a^2)^2 > 0$, but we can have $A<0$ for some $\kappa<1$.  Also for $\kappa=1$ the solutions in \rf{1.10} reduce to those given in \rf{4.19}, with \rf{1.6} given by \rf{4.20}.

\section{Reaction diffusion systems and derivation of the metric in \rf{1.6}}

Since the metric \rf{1.6} is one of the main objects of interest we indicate in this section its derivation.  For a constant $B$ consider the system of partial differential equations
\beqa
r_t - r_{xx} + Br^2s &=& 0\label{2.1}\\
s_t + s_{xx} - Brs^2 &=& 0\nonumber
\eeqa
in the variables $(x,t)$.  This system is a special case of the more general reaction diffusion system (RDS)
\beqa
r_t &=& d_r r_{xx} + F(r,s)\label{2.2}\\
s_t &=& d_s s_{xx} + G(r,s)\nonumber
\eeqa
that occurs in chemistry, physics, or biology, for example, where $d_r, d_s$ are diffusion constants, and $F,G$ are growth and interaction functions.  The key point for us is that from solutions $r(x,t), s(x,t)$ of \rf{2.1} one can construct a metric $g$ of constant Ricci scalar curvature $R(g) = 4B$ by the following prescription \cite{5},\cite{6},\cite{7}:
\beqa
g_{11}\SR{def.}{=} -r_x s_x,&& \qquad g_{12} \SR{def.}{=} \frac{1}{2}\left( sr_x - rs_x \right), \qquad g_{22}\SR{def.}{=} rs \nonumber\\
&& ds^2 \SR{def.}{=} g_{11}dt^2 + 2g_{12}dtdx + g_{22}dx^2.\label{2.3}
\eeqa
One could also simply start with the definitions in \rf{2.3}, apart from the preceding references that employ Cartan's zweibein formalism \cite{8}, and use a Maple program (tensor), for example, to check directly that indeed $R(g) = 4B$.  Our interest is in the choice $B = 2/b^2$, where for real $a,b,v$, with $a,b\neq 0$ as in section 1, $r(x,t), s(x,t)$ given by 
\beqa
r(x,t) \SR{def.}{=} ab\dnaxk {\rm exp}\left({\left[ \frac{v^2}{4} + a^2(2-\kappa^2)\right]t - \frac{vx}{2}}\right)\label{2.4}\\
s(x,t) \SR{def.}{=} -ab\dnaxk {\rm exp}\left({ - \left[ \frac{v^2}{4} + a^2(2-\kappa^2)\right]t + \frac{vx}{2}}\right)\nonumber
\eeqa
are solutions the system \rf{2.1}, which also could be checked directly by Maple.  For $B = 2/b^2$, \rf{2.1} is system \rf{1.8} with solutions \rf{2.4} promised in section 1, and $g$ in \rf{2.3} has the scalar curvature $4B = 8/b^2$ discussed in section 1.  From \cite{4} various formulas like
\beqa
\snSxk  + \cnSxk  = 1, \qquad \dnSxk  + \kappa^2 \snSxk  = 1\nonumber\\
\frac{d}{dx}\snxk = \cnxk \dnxk,  \quad  \frac{d}{dx}\cnxk  = -\snxk \dnxk \nonumber\\
\qquad\qquad \frac{d}{dx}\dnxk  = -\kappa^2 \snxk \cnxk  \label{2.5}
\eeqa
are available.  Using the prescription \rf{2.3} one computes that 
\beqa
g_{11}&=& a^2 b^2 \left[ a^2 \kappa^4 \snSaxk   \cnSaxk -\frac{v^2}{4}\dnSaxk \right]\nonumber\\
g_{12}&=&\frac{a^2b^2v}{2}\dnSaxk , \quad g_{22}=-a^2b^2\dnSaxk .\label{2.6}
\eeqa
For $\rho\SR{def.}{=}a(x-vt)$, so that $d\rho = a(dx-vdt)$, $g$ can be expressed more conveniently as 
\beq
 ds^2 = a^2b^2\dnSrk \left[ \left( \frac{a^2\kappa^4 \snSrk \cnSrk }{\dnSrk } - \frac{v^2}{4}\right)dt^2 - \frac{v}{a}dtd\rho - \frac{d\rho^2}{a^2}\right].\label{2.7}
\eeq

The goal now is to set up a change of variables $(t,\rho)\longrightarrow(\tau,\rho)$ so that $g$ in \rf{2.7} is transformed to \rf{1.6} -- where the cross term $d\tau d\rho$ does \emph{not} appear, in comparison with the term $dtd\rho$ appearing in \rf{2.7}.  For this purpose note first, in general, that for 
\beq
h = A(\rho) dt^2 + C_1(\rho)d\rho dt + C_2(\rho)d\rho^2\label{2.8}
\eeq
the change of variables $\tau = t + \phi(\rho)$ gives $dt = d\tau - \phi'(\rho) d\rho$, $dt^2 = d\tau^2 - 2\phi'(\rho) d\tau d\rho + \phi'(\rho)^2 d\rho^2$, and 
\beq
 h = A(\rho)d\tau^2 + \left[ -2\phi'(\rho)A(\rho) + C_1(\rho)\right]d\tau d\rho + \left[A(\rho)\phi'(\rho)^2 - C_1(\rho)\phi'(\rho) + C_2(\rho)\right] d\rho^2.\label{2.9}
\eeq
The condition that the cross term $d\tau d\rho$ does not appear is therefore that $\phi(\rho)$ satisifies
\beq
\phi'(\rho) = \frac{C_1(\rho)}{2A(\rho)}.\label{2.10}
\eeq

Apply this to \rf{2.7}:
\beq
\phi'(\rho) = \frac{-v}{2a}\cdot \left( \frac{a^2 \kappa^4 \snSrk \cnSrk }{\dnSrk } - \frac{v^2}{4}\right)^{-1}.\label{2.11}
\eeq
Now by \rf{2.5}, $\dnSxk  = 1-\kappa^2 \snSxk  = \snSxk  + \cnSxk  - \kappa^2 \snSxk  = \cnSxk  + (1-\kappa^2)\snSxk \geq \cnSxk $ and $\snSxk  \leq \snSxk  + \cnSxk  = 1$ $\Rightarrow$ $\snSxk \cnSxk /\dnSxk  \leq 1$.  If the term in parenthesis in \rf{2.11} were zero, this would therefore force the inequality $v^2/4a^2\kappa^4 \leq 1$.  That is, if $\left| v/2a\kappa^2\right| > 1$, which is the assumption in \rf{1.7}, then $v^2/4a^2\kappa^4 > 1$ and therefore the denominator term in parenthesis in \rf{2.11} is non-zero, which means that $\phi'(\rho)$ is a continuous function and \rf{2.11} therefore indeed has a solution $\phi(\rho)$, with the assumption \rf{1.7} imposed.  Also, the coefficient of $d\rho^2$ in \rf{2.9} is
\beq
a^2b^2\dn^2 Q \frac{v^2}{4a^2Q^2} - a b^2v\dn^2 \frac{v}{2aQ} - b^2\dn^2 = -\frac{b^2v^2\dn^2}{4Q} - b^2\dn^2\label{2.12}
\eeq
where for convenience we write sn,cn,dn for $\snrk,\cnrk,\dnrk$ and $Q$ for 
\beq
Q(\rho,\kappa) \SR{def.}{=}\frac{a^2\kappa^4 \snSrk \cnSrk }{\dnSrk}  - \frac{v^2}{4}.\label{2.13}
\eeq
Then
\beqa
&& Q + \frac{v^2}{4} = \frac{a^2\kappa^4 \sn^2 \cn^2}{\dn^2} \Rightarrow \nonumber\\
&&1 + \frac{v^2}{4Q} = \frac{a^2\kappa^4 \sn^2\cn^2}{\dn^2 Q} \Rightarrow\label{2.14}\\
&&-b^2\dn^2 - \frac{b^2v^2\dn^2}{4Q} = -\frac{a^2b^2\kappa^4\sn^2\cn^2}{Q},\nonumber
\eeqa
which is the coefficient of $d\rho^2$ in \rf{2.9} by \rf{2.12}.  Then by  \rf{2.10},\rf{2.9} reads 
\beq
h = a^2b^2\dn^2Qd\tau^2 - \frac{a^2b^2\kappa^4\sn^2\cn^2}{Q}d\rho^2,\label{2.15}
\eeq
which is \rf{1.6}.  That is, we have verified that the change of variables $\tau = t + \phi(\rho)$ with $\phi(\rho)$ subject to condition \rf{2.11} (which in fact renders $\phi'(\rho)$ a continuous function, again assuming \rf{1.7}) transforms the reaction diffusion metric in \rf{2.7} to the diagonal metric in \rf{1.6}.

In the special case when the elliptic modulus $\kappa=1$
\beq
\sn(x,1) = \tanh(x), \quad \cn(x,1) = \dn(x,1) = \sech(x)\label{2.16}
\eeq
and \rf{2.7}, \rf{1.6} simplify:
\beqa
 ds^2 &=& a^2b^2 \sech^2\rho \left[ \left( a^2 \tanh^2\rho -\frac{v^2}{4}\right)dt^2 - \frac{v}{a}dtd\rho - \frac{d\rho^2}{a^2} \right],\label{2.17}\\
 ds^2 &=& a^2b^2\sech^2\rho\left[ \left( a^2 \tanh^2\rho - \frac{v^2}{4}\right)dt^2 - \tanh^2\rho \left( a^2\tanh^2\rho - \frac{v^2}{4}\right)^{-1}d\rho^2 \right],\nonumber
\eeqa
which are the line elements (3.12), (3.14), respectively, in \cite{6}; $a$ here corresponds to the notation $k$ there.  Also the cosmological constant $\Lambda_0$ in \cite{6} corresponds to our $2\Lambda = -8/b^2$ : $b^2 = 8/(-\Lambda_0)$.  Similarly, $r$ and $s$ in \rf{2.4} reduce to the dissipative soliton solutions $q^+$ and $q^-$, respectively, in (2.32) of \cite{6}, apart from the factor $b$.  One can also explicitly determine $\phi(\rho)$ in \rf{2.11}.

\section{Transformation of the metric in \rf{1.6} to a J-T black hole metric}

Now that the existence of the metric in \rf{1.6} has been  described in the context of a reaction diffusion system (namely \rf{1.8}), the strategy of this section is to set up a series of changes of variables, as indicated in the introduction, that transforms it to the simpler J-T form \rf{1.4}.  Other applications, of independent interest, can flow from this -- apart from our main focus to solve the system \rf{1.9}.  A general method to go from \rf{1.6} to \rf{1.4} has been developed by the first named author.  Alternatively, one can generalize part of the argument in \cite{6} that leads at least to a Schwarzschild form, as we do here, and then argue a bit more to obtain the J-T form -- the final result being expressed by equations \rf{3.11}-\rf{3.13} below.

Start with the change of variables $r=|a|\dnrk$ so that $dr = -\kappa^2|a|\snrk\cdot$ $\cnrk d\rho$ by \rf{2.5} $\Rightarrow$
\beq
\frac{dr^2}{r^2} = \frac{\kappa^4\snSrk \cnSrk d\rho^2}{\dnSrk}.\label{3.1}
\eeq
Also by \rf{2.5}

\beqa
&&\frac{a^2\kappa^4 \snSrk \cnSrk}{\dnSrk} = \frac{a^2 \kappa^2 (1-\dnSrk)(1-\snSrk)}{\dnSrk}\nonumber\\
&& = \frac{(a^2 - a^2\dnSrk)(\kappa^2 - \kappa^2 \snSrk)}{\dnSrk}\nonumber\\
&& = \frac{(a^2 - r^2)(\kappa^2 + \dnSrk - 1)}{r^2/a^2} = \frac{(a^2 - r^2)(\kappa^2 - 1 + r^2/a^2)}{r^2/a^2}\label{3.2}\\
&& = \frac{(a^2-r^2)}{r^2}\left[ a^2(\kappa^2 - 1) + r^2\right] \Rightarrow \nonumber\\
&& \frac{a^2\kappa^4\snSrk \cnSrk}{\dnSrk} - \frac{v^2}{4} = \nonumber\\
&&\frac{(a^2-r^2)}{r^2}\left[ a^2(\kappa^2 - 1) + r^2\right]  - \frac{v^2}{4} =\nonumber\\
&& \frac{a^4(\kappa^2-1)}{r^2} + 2a^2 - a^2\kappa^2 - r^2 - \frac{v^2}{4} = \nonumber\\
&& \frac{a^4(\kappa^2-1)}{r^2} + 2a^2 - a^2\kappa^2 - r^2  - r_0^2-a^2\kappa^4\nonumber
\eeqa
for 
\beq
r_0^2 \SR{def.}{=} \frac{v^2}{4} - a^2\kappa^4 > 0.\label{3.3}
\eeq
Again by \rf{1.7}, $v^2/4a^2\kappa^4 > 1 \Rightarrow v^2/4 > a^2\kappa^4 \Rightarrow$ indeed $r_0^2 > 0$.  By \rf{3.1}, \rf{3.2} we see that we can write \rf{1.6} as 
\beqa
 g= b^2 \left[ -r^2\left( r^2 + r_0^2 + a^2(\kappa^4 + \kappa^2 - 2) + \frac{a^4(1-\kappa^2)}{r^2}\right)d\tau^2\right. \nonumber\\
 \left.+ \left( r^2 + r_0^2 + a^2(\kappa^4 + \kappa^2 - 2) + \frac{a^4(1-\kappa^2)}{r^2}\right)^{-1} dr^2\right].\label{3.4}
\eeqa

Next let $x\SR{def.}{=}(2r^2 + r_0^2)/r_0^4$, as in (3.18) of \cite{6} but where our $r_0^2$ in \rf{3.3} generalizes their $r_0^2$, and for convenience let 
\beq
\alpha\SR{def.}{=}a^2(\kappa^4 + \kappa^2 - 2), \quad \beta\SR{def.}{=}a^4(1-\kappa^2)\label{3.5}
\eeq
in \rf{3.4}.  Then $g$ in \rf{3.4} assumes the form
\beqa
 g= b^2 \left[ \left( -\frac{r_0^4}{4}(r_0^4x^2 - 1) - \frac{\alpha r_0^2}{2}(r_0^2x-1)-\beta\right) d\tau^2\right. \nonumber\\
  \left. + \frac{1}{16}r_0^8 \left( \frac{r_0^4}{4}(r_0^4x^2-1) + \frac{\alpha}{2}r_0^2(r_0^2x-1)+\beta\right)^{-1}  dx^2\right]\nonumber
  \eeqa
  \beqa
 = \frac{b^2r_0^4}{4}\left[ - \left( r_0^4 x^2 - 1 + \frac{2\alpha}{r_0^2}(r_0^2x-1) + \frac{4\beta}{r_0^4}\right)d\tau^2\right. \nonumber\\
 \left. + \left(r_0^4x^2-1 + \frac{2\alpha}{r_0^2}(r_0^2x-1) + \frac{4\beta}{r_0^4}\right)^{-1}dx^2\right]\label{3.6}
\eeqa
which generalizes the Schwarzchild form (3.19) of \cite{6} since for $\kappa=1$ we have that $\alpha = \beta = 0$ in \rf{3.5}. 

For the change of variables $t=A_0\tau$, $r_-=A_0x$ with $A_0\SR{def.}{=}|b|r_0^2/2$, the Schwarzschild $g$ in \rf{3.6} goes to
\beqa
 g=-\left[ \frac{4}{b^2}r_-^2-1 + \frac{2\alpha}{r_0^2}\left( \frac{2}{|b|}r_- - 1\right) + \frac{4\beta}{r_0^4}\right]dt^2\nonumber\\
 + \left[\frac{4}{b^2}r_-^2 - 1 + \frac{2\alpha}{r_0^2}\left( \frac{2}{|b|}r_- -1\right) + \frac{4\beta}{r_0^4}\right]^{-1}dr_-^2 , \label{3.7}
\eeqa
which in turn goes to
\beqa
 g= -\left[ \frac{4}{b^2}r_1^2 - b^2 + \frac{2\alpha}{r_0^2}\left( \frac{2b}{|b|} r_1 - b^2\right) + \frac{4\beta b^2}{r_0^4}\right]dT^2 \label{3.8}\\
 + \left[ \frac{4r_1^2}{b^2} - b^2 + \frac{2\alpha}{r_0^2}\left( \frac{2b}{|b|} r_1 - b^2\right) + \frac{4\beta b^2}{r_0^4}\right]^{-1}dr_1^2\nonumber
\eeqa
by way of the change of variables $t=bT$, $r_- = r_1/b$.  We need one final observation:  In general a metric of the form
\beq
g_1 = -\left[ A_1x^2 + B_1x + C_1 \right] dT^2 + \left[A_1x^2 + B_1x + C_1\right]^{-1}dx^2,\label{3.9}
\eeq
say $A_1\neq 0$ can be transformed to the J-T form \rf{1.4}, namely
\beq
g_1 = -\left[ A_1r^2 + C_1 - \frac{B_1^2}{4A_1}\right]dT^2 + \left[A_1r^2 + C_1 - \frac{B_1^2}{4A_1}\right]^{-1}dr^2,\label{3.10}
\eeq
by way of the change of variables $r=x+ \frac{B_1}{2A_1}$.  Apply this to \rf{3.8} with $x$ playing the role of $r_1$ there:
\beq
g = -\left[ A_1r^2 + C_1 - \frac{B_1^2}{4A_1}\right]dT^2 + \left[A_1r^2 + C_1 - \frac{B_1^2}{4A_1}\right]^{-1}dr^2,\label{3.11}
\eeq
for 
\beq
A_1\SR{def.}{=}\frac{4}{b^2}, \quad B_1\SR{def.}{=}\frac{4\alpha b}{r_0^2 |b|}, \quad C_1 \SR{def.}{=}-b^2 - \frac{2\alpha b^2}{r_0^2} + \frac{4\beta b^2}{r_0^4}.\label{3.12}
\eeq
Using definition \rf{3.5} for $\alpha,\beta$ and $r_0^2 = \frac{v^2}{4} - a^2 \kappa^4$, which is definition \rf{3.3}, one computes that
\beq
C_1 - \frac{B_1^2}{4A_1} = \frac{-b^2}{r_0^4}\left[ \frac{v^4}{16} - \frac{a^2v^2}{2}(2-\kappa^2) + a^4 \kappa^4\right] \label{3.13}
\eeq
in \rf{3.11}.

\section{Derivation of the solutions \rf{1.10} of the field equations \rf{1.9}}

The main result is derived in this section.  Namely, we indicate how the series of changes of variables in section 3 (according to remarks in the introduction) lead to the linearly independent solutions $\Phi^{(j)}(\tau,\rho)$, $j=1,2,3$ in \rf{1.10} of the dilaton field equations in \rf{1.9}.  There the metric elements $g_{ij}$ are given by \rf{1.6}:  For $Q(\rho,\kappa)$ in \rf{2.13}
\beqa
g_{11}&\SR{def.}{=}& a^2 b^2 \dnSrk Q(\rho,\kappa), \quad g_{12}=g_{21}=0,\nonumber\\
g_{22}&=& a^2 b^2 \dnSrk \left( \frac{-\kappa^4 \snSrk \cnSrk}{\dnSrk}\right)Q(\rho,\kappa)^{-1},\label{4.1}
\eeqa
and the $\nabla_i \nabla_j \Phi$ are given by \rf{1.3} for $(x_1,x_2) = (\tau,\rho)$.  The Christoffel symbols $\Gamma_{ij}^k$ in \rf{1.3} (which could be computed by Maple, for example) will not be needed for the derivation of \rf{1.10}, although they could be used to verify these solutions.  Obviously any dilaton solution could be replaced by any non-zero multiple of itself.  In the following then, we can disregard such multiples if we wish to.

In addition to the dilaton solution $\Phi^{(1)}(T,r)\SR{def.}{=}mr$ in \rf{1.5} for the metric \rf{1.4} in the variables $(T,r)$, there are solutions
\beqa
\Phi^{(2)}(T,r)&\SR{def.}{=}&\sqrt{m^2r^2-M}\sinh(m\sqrt{M}T)\nonumber\\
\Phi^{(3)}(T,r)&\SR{def.}{=}&\sqrt{m^2r^2-M}\cosh(m\sqrt{M}T).\label{4.2}
\eeqa
We work backwards the changes of variables in section 3 for $\Phi^{(1)}(T,r), \Phi^{(2)}(T,r)$, for example, to see how one arrives at the first two solutions $\Phi^{(1)}(\tau,\rho),\Phi^{(2)}(\tau,\rho)$ in \rf{1.10}, in the variables $(\tau,\rho)$.

Starting with the \rf{3.11} version of \rf{1.4}, we have $m^2 = A_1 = 4/b^2$ by \rf{3.12}, with $M=-(C_1 - B_1^2/4A_1)$ given by \rf{3.13}.  Here $m\sqrt{M} = \sqrt{B_1^2 - 4A_1C_1}/2$ (for $m=2/|b|$) $\Rightarrow$
\beq
\Phi^{(1)}(T,r) = \frac{2}{|b|}r, \quad \Phi^{(2)}(T,r)=\sqrt{A_1r^2 + C_1 - \frac{B_1^2}{4A_1}}\sinh\left( \frac{\sqrt{B_1^2-4A_1C_1}}{2}T\right).\label{4.3}
\eeq
By the final change of variables $r=r_1 + \frac{B_1}{2A_1}$ in section 3, we see that $A_1r^2 = A_1r_1^2 + B_1r_1 + B_1^2/4A_1$ $\Rightarrow$
\beqa
\Phi^{(1)}(T,r_1) &=& \frac{2}{|b|}\left( r_1 + \frac{B_1}{2A_1}\right),\label{4.4}\\
\Phi^{(2)}(T,r_1) &=& \sqrt{A_1r_1^2 + B_1r_1 + C_1}\sinh\left(\frac{\sqrt{B_1^2-4A_1C_1}}{2}T\right).\nonumber
\eeqa
The change of variables $t=bT$, $r_-=r_1/b$ preceded the change $r=r_1 + B_1/2A_1$, so that 
\beqa
\Phi^{(1)}(t,r_-) &=& \frac{2b}{|b|}r_- + \frac{B_1}{|b|A_1},\label{4.5}\\
\Phi^{(2)}(t,r_-) &=&\sqrt{4r_-^2 + B_1 br_- + C_1}\sinh\left(\frac{\sqrt{B_1^2-4A_1C_1}}{2b}t\right)
\nonumber
\eeqa
since $A_1b^2 \SR{def.}{=}4$.  We had $t=A_0\tau$, $r_-=A_0x$ for $A_0\SR{def.}{=}|b|r_0^2/2$, which gives
\beqa
\Phi^{(1)}(\tau,x) &=& br_0^2x +  \frac{B_1}{|b|A_1}, \quad \frac{|b|}{b}\Phi^{(1)}(\tau,x) = |b|r_0^2x + \frac{B_1}{bA_1},\label{4.6}\\
\Phi^{(2)}(\tau,x)&=& \sqrt{b^2r_0^4x^2 + \frac{B_1 b|b|}{2} r_0^2 x + C_1}\sinh\left( \frac{\sqrt{B_1^2-4A_1C_1}}{4} \cdot \frac{|b|r_0^2}{b} \tau\right),\nonumber
\eeqa
for the Schwarzchild version of our metric in \rf{3.6}.  Next let $x=(2r^2 + r_0^2)/r_0^4$ to get 
\beqa
 \Phi^{(1)}(\tau,r) &=&|b|(2r^2 + r_0^2)/r_0^2 + \frac{B_1}{bA_1},\label{4.7}\\
 \Phi^{(2)}(\tau,r) &=&\sqrt{  \frac{b^2(2r^2 + r_0^2)^2}{r_0^4} + \frac{B_1 b|b|}{2} \frac{(2r^2 + r_0^2)}{r_0^2} + C_1}\sinh\left( \frac{\sqrt{B_1^2-4A_1C_1}}{4}r_0^2 \tau\right)\nonumber
\eeqa
where we have disregarded the multiple $|b|/b = \pm 1$ in \rf{4.6} and have used that $\sinh(|b|x/b) = (|b|/b)\sinh(x)$.  Finally, the first change of variables $r=|a|\dnrk$ in section 3 gives
\beq
\Phi^{(1)}(\tau,\rho) = \frac{|b|}{r_0^2} \left( 2a^2 \dnSrk + r_0^2 \right) + \frac{\alpha |b|}{r_0^2},
\label{4.8}
\eeq
by definition \rf{3.12}.  If we disregard the multiple $|b|/r_0^2$ in \rf{4.8} and use that $r_0^2 + \alpha \SR{def.}{=} v^2/4 + a^2(\kappa^2-2)$ by definitions \rf{3.3}, \rf{3.5} we obtain from \rf{4.8} the first solution 
\beq
\Phi^{(1)}(\tau,\rho) = 2a^2 \dnSrk + \frac{v^2}{4} + a^2(\kappa^2-2)\label{4.9}
\eeq
in \rf{1.10}.  More work is required of course to obtain the second solution there.

First we note that by \rf{3.12}, \rf{3.13}
\beqa
B_1^2 - 4A_1C_1=-4A_1\left( C_1 - \frac{B_1^2}{4A_1}\right) = \frac{16}{r_0^4}\left[ \frac{v^4}{16} - \frac{a^2 v^2}{2}(2-\kappa^2) + a^4 \kappa^4\right] \Rightarrow \nonumber\\
\frac{\sqrt{B_1^2 - 4A_1C_1}}{4}r_0^2\tau = \sqrt{\frac{v^4}{16} - \frac{a^2 v^2}{2}(2-\kappa^2) + a^4 \kappa^4}\cdot\tau,\label{4.10}
\eeqa
which is the $\sqrt{A}\tau$ in \rf{1.10}.  Also for $r=|a|\dn$, $\dn=\dnrk$, the quantity under the other radical in \rf{4.7} is
\beqa
\frac{b^2}{r_0^4}\left( 2a^2 \dn^2 + r_0^2\right)^2 + \frac{B_1 b|b|}{2r_0^2}(2a^2\dn^2 + r_0^2) + C_1=\label{4.11}\\
\frac{4a^4b^2}{r_0^4}\dn^4 + \left( \frac{4a^2 b^2}{r_0^2} + \frac{B_1 b|b| a^2}{r_0^2}\right)\dn^2+b^2 + \frac{B_1 b|b|}{2} + C_1,\nonumber
\eeqa
where by definition \rf{3.12}
\beqa
\frac{B_1 b|b| a^2}{r_0^2} = \frac{4a^2b^2 \alpha}{r_0^4}, \quad b^2 + \frac{B_1 b|b|}{2} + C_1 =\nonumber\\
b^2 + \frac{2b^2\alpha}{r_0^2} - b^2 - \frac{2\alpha b^2}{r_0^2} + \frac{4\beta b^2}{r_0^4} = \frac{4\beta b^2}{r_0^4} \Rightarrow \label{4.12}\\
\frac{4a^2 b^2}{r_0^2} + \frac{B_1 b|b|a^2}{r_0^2} = \frac{4r_0^2 a^2 b^2 + 4a^2 b^2 \alpha}{r_0^4} = \frac{4a^2b^2}{r_0^4}(r_0^2+\alpha)\nonumber\\
=\frac{4a^2 b^2}{r_0^4}\left[ \frac{v^2}{4} + a^2(\kappa^2 - 2)\right],\nonumber
\eeqa
again by definitions \rf{3.3}, \rf{3.5}.  That is, since $\beta = a^4 (1-\kappa^2)$ by definition \rf{3.5} the quantity in \rf{4.11} (which is under the radical in \rf{4.7} for $r=|a|\dn$) is 
\beqa
\frac{4a^4 b^2}{r_0^4 }\dn^4 + \frac{4a^2b^2}{r_0^4}\left[ \frac{v^2}{4} + a^2(\kappa^2 - 2)\right]\dn^2 + \frac{4a^4b^2(1-\kappa^2)}{r_0^4} = \label{4.13}\\
\frac{4a^2 b^2}{r_0^4}\left[ a^2 \dn^4 + \left(\frac{v^2}{4} + a^2(\kappa^2-2)\right)\dn^2 + a^2(1-\kappa^2)\right].\nonumber
\eeqa
We let $B(\rho)$ denote the latter bracket here.  By \rf{4.7}, \rf{4.10}, \rf{4.13} we see that (for now) $\Phi^{(2)}(\tau,\rho) = \sqrt{B(\rho)}\sinh\left( \sqrt{A}\tau\right)$, if we disregard the multiple $\sqrt{4a^2b^2/r_0^4} = 2|a||b|/r_0^2$. 

We find an alternate expression for $B(\rho)$, which is simpler and which shows that $B(\rho)>0$, given \rf{1.7}.  Again we write sn, cn, dn for $\snrk, \cnrk,\dnrk$, and we make use of \rf{2.5}.
\beqa
B(\rho)&\SR{def.}{=}&\dn^2\left[ a^2 \dn^2 + \frac{v^2}{4} - 2a^2 + a^2\kappa^2 + a^2 \frac{(1-\kappa^2)}{\dn^2}\right]\nonumber\\
&=&\dn^2\left[ a^2(1-\kappa^2 \sn^2) + \frac{v^2}{4} - 2a^2 + a^2\kappa^2 + a^2 \frac{(1-\kappa^2)}{\dn^2}\right]\nonumber\\
&=&\dn^2\left[ -a^2\kappa^2 \sn^2 + \frac{v^2}{4} - a^2(1-\kappa^2) + a^2 \frac{(1-\kappa^2)}{\dn^2}\right]\nonumber\\
&=&\dn^2\left[ \frac{v^2}{4} - a^2\kappa^2\sn^2 + a^2 \frac{(1-\kappa^2)}{\dn^2}(1-\dn^2)\right]\label{4.14}\\
&=&\dn^2\left[ \frac{v^2}{4} - a^2\kappa^2\sn^2 + a^2 \frac{(1-\kappa^2)}{\dn^2}\kappa^2\sn^2\right]\nonumber\\
&=&\dn^2\left[ \frac{v^2}{4} - a^2\kappa^2\sn^2 \frac{(\dn^2-1+\kappa^2)}{\dn^2}\right]\nonumber\\
&=&\dn^2\left[ \frac{v^2}{4} - \frac{a^2\kappa^2\sn^2}{\dn^2} (-\kappa^2\sn^2+\kappa^2)\right]\nonumber\\
&=&\dn^2\left[ \frac{v^2}{4} - \frac{a^2\kappa^4\sn^2}{\dn^2} (1-\sn^2)\right]\nonumber\\
&=&\dn^2\left[ \frac{v^2}{4} - \frac{a^2\kappa^4\sn^2\cn^2}{\dn^2} \right],\nonumber
\eeqa
where we noted in section 2 that $\sn^2\cn^2/\dn^2\leq 1$.  Hence
\beq
\frac{v^2}{4} - a^2\kappa^4 \frac{\sn^2\cn^2}{\dn^2} \geq \frac{v^2}{4} - a^2\kappa^4 > 0
\label{4.15}
\eeq
by \rf{1.7}, again as in \rf{3.3}, and we see that $B(\rho) > 0$ since $\dnrk \neq 0$ for $\rho$ a real number.  Moreover we have established the desired expression for $\Phi^{(2)}(\tau,\rho)$ in \rf{1.10}.  Clearly one can replace the hyperbolic sine in the preceding discussion by the hyperbolic cosine in \rf{4.2} to obtain the third solution $\Phi^{(3)}(\tau,\rho)$ in \rf{1.10}.  To finish other claims made in section 1 we check that in \rf{1.11} $A=0$ only for $a=\pm (1-\sqrt{1-\kappa^2})v/2\kappa^2$.  We continue to assume \rf{1.7} of course.  

The quartic equation $A=0$ has roots $a=\pm(1+\sqrt{1-\kappa^2})v/2\kappa^2$, $\pm(1-\sqrt{1-\kappa^2})v/2\kappa^2$ with $a^2 = (2-\kappa^2 + 2\sqrt{1-\kappa^2})v^2/4\kappa^4$, $(2-\kappa^2-2\sqrt{1-\kappa^2})v^2/4\kappa^4$ respectively.  \rf{1.7} requires that $a^2 < v^2/4\kappa^4$, which forces the inequalities
\beq
2-\kappa^2 + 2\sqrt{1-\kappa^2} < 1, \quad 2-\kappa^2 - 2\sqrt{1-\kappa^2} < 1,\label{4.16}
\eeq
of which the first one reads $1-\kappa^2 + 2\sqrt{1-\kappa^2}<0$, with the left hand side here being $\geq 0$ -- a contradiction.  That is, we cannot have $a=\pm (1+\sqrt{1-\kappa^2})v/2\kappa^2$ which means that $a=\pm (1-\sqrt{1-\kappa^2})v/2\kappa^2$.  Also we check that the solutions are linearly independent:  Assume for constants $c_1,c_2,c_3$ that
\beq
c_1\Phi^{(1)}(\tau,\rho) + c_2\Phi^{(2)}(\tau,\rho) + c_3\Phi^{(3)}(\tau,\rho) = 0.
\label{4.17}
\eeq
Differentiate this equation with respect to $\tau$ and evaluate the result at $(\tau,0)$:
\beq
c_2\sqrt{A} \cosh(\sqrt{A}\tau) |v|/2 + c_3 \sqrt{A}\sinh(\sqrt{A}\tau)|v|/2 = 0
\label{4.18}
\eeq
since $\dn(0,\kappa)=1, \sn(0,\kappa)=0$.  The choice $\tau=0$ then gives $c_2=0$, since $A,v\neq 0$, and differentiation of the equation $c_3\sqrt{A}\sinh(\sqrt{A}\tau)|v|/2 = 0$ and at $\tau=0$ gives $c_3=0$.  Using again that $\dn(0,\kappa)=1$ we see by \rf{1.10} that $\Phi^{(1)}(\tau,0) \SR{def.}{=} v^2/4 + a^2\kappa^2>0$ and hence also $c_1=0$.

Note that if $v=a=2$ and $\kappa=1/2$, for example, then even though $v/2a\kappa^2 = 2>1$ (so that \rf{1.7} is satisfied), we have that $A=-12<0$. 

Again in the special case when the elliptic modulus $\kappa=1$, we have in \rf{1.11} that $A=(v^2/4-a^2)^2  > 0$ and $\sqrt{A} = v^2/4 - a^2 > 0$  (by \rf{1.7} or \rf{3.3}), and $B(\rho) = \sech^2\rho \left[ v^2/4 - a^2\tanh^2\rho\right] = \sech^2\rho \left[ v^2 - 4a^2 \tanh^2\rho\right]/4$, by \rf{2.16}, \rf{4.14} $\Rightarrow$ $\sqrt{B(\rho)} = \frac{1}{2}\sech\rho \sqrt{v^2 - 4a^2\tanh^2\rho}.$  Here (directly) $\tanh^2\rho \leq 1$ $\Rightarrow$ $v^2 - 4a^2 \tanh^2\rho > 0$, again as $v^2 > 4a^2$.  Thus by \rf{1.10} and \rf{2.17}
\beqa
\Phi^{(1)}(\tau,\rho) &=& 2a^2 \sech^2\rho + \frac{v^2}{4} - a^2,\nonumber\\
\Phi^{(2)}(\tau,\rho) &=& \left(\sinh \left( \frac{v^2}{4} - a^2\right)\tau\right) \left(\sech\rho\right) \sqrt{v^2 - 4a^2\tanh^2\rho}, \label{4.19}\\
\Phi^{(3)}(\tau,\rho) &=& \left(\cosh \left( \frac{v^2}{4} - a^2\right)\tau\right) \left(\sech\rho\right) \sqrt{v^2 - 4a^2\tanh^2\rho}\nonumber
\eeqa
(where we have disregarded the multiple $1/2$ in $\sqrt{B(\rho)}$) are dilaton field solutions for the metric
\beq
 ds^2 = a^2b^2 \sech^2\rho\left[ \left( a^2\tanh^2\rho  - \frac{v^2}{4}\right) d\tau^2 - \left(\tanh^2\rho\right) \left( a^2 \tanh^2 \rho - \frac{v^2}{4}\right)^{-1} d\rho^2 \right].\label{4.20}
\eeq
The solutions in \rf{4.19} are also new.

\section{Killing vector fields for the solutions \rf{1.6}, \rf{1.10}}

Recall that a smooth vector field $Y$ on an $n$-dimensional Riemannian manifold $(M,g)$ is called a \emph{Killing} vector field (or an infinitesimal motion of $M$) if for arbitrary smooth vector fields $X,Z$ on $M$
\beq
Yg(X,Z) = g\left([Y,X],Z\right) + g(X,[Y,Z])=0.\label{5.1}
\eeq
If $Y = \displaystyle\sum_{i=1}^n Y_i \frac{\partial}{\partial x_i}$ is an expression of $Y$ in terms of local coordinates $(x_1, \dots, x_n)$ on $M$, then \rf{5.1} is equivalent to the system of equations
\beq
\displaystyle\sum_{i=1}^n g_{ki}\frac{\partial Y_i}{\partial x_j} + g_{ji}\frac{\partial Y_i}{\partial x_k} + \frac{\partial g_{jk}}{\partial x_i} Y_i = 0 \label{5.2}
\eeq
for $1\leq j,k\leq n$ \cite{8}, \cite{9}.  In the special (diagonal) case with $g_{ij}=0$ for $i\neq j$, and with $n=2$, the Killing equations \rf{5.2} simplify to the following three equations:
\beqa
2g_{11}\frac{\partial Y_1}{\partial x_1} + \frac{\partial g_{11}}{\partial x_1}Y_1 + \frac{\partial g_{11}}{\partial x_2} Y_2 &=& 0\nonumber\\
\qquad \qquad  g_{11} \frac{\partial Y_1}{\partial x_2} + g_{22}\frac{\partial Y_2}{\partial x_1} &=& 0\label{5.3}\\
\frac{\partial g_{22}}{\partial x_1}Y_1 + 2g_{22}\frac{\partial Y_2}{\partial x_2} + \frac{\partial g_{22}}{\partial x_2}Y_2 &=& 0\nonumber
\eeqa

As was shown in \cite{10}, every solution $(g,\Phi)$ of the field equations in \rf{1.1} gives rise to a corresponding Killing vector field $Y=Y(g,\Phi)$ by way of the local prescription
\beq
Y_i = \frac{l \epsilon^{ij}}{\sqrt{|\mbox{det}g|}} \frac{\partial \Phi}{\partial x_j}\label{5.4}
\eeq
with $\epsilon^{ij}$ = a permutation symbol.  $Y$ preserves both $g$ and $\Phi$.  For $g$ in \rf{1.4} and for the fields $\Phi$ in \rf{1.5} and \rf{4.2}, the corresponding Killing vector fields are given in equations (16), (17), (18) of \cite{11}, for example.  Our interest of course is in the case of the three solutions $(g,\Phi^{(j)})$ in \rf{1.10} with $g$ given by \rf{1.6}.  By \rf{4.1}, $\sqrt{|\mbox{det}g|} = a^2 b^2 \kappa^2 | \sn \cn | \dn$.  Since $Y_i$ could be replaced by a scalar multiple of itself (for example, $-Y_i$), we shall disregard the absolute value of $\sn\cn$ here, and given \rf{1.9} we shall take $l = \frac{b}{2}$ (instead of $\frac{|b|}{2}$).  For $\epsilon^{11} = \epsilon^{22} = 0, \epsilon^{12} = -1 = -\epsilon^{21}$ \rf{5.4} then assumes the generic form 
\beqa
Y_1 &=& \left[ 2a^2 b \kappa^2 \snrk \cnrk \dnrk\right]^{-1} \left( -\frac{\partial \Phi}{\partial \rho} \right)\label{5.5}\\
Y_2  &=& \left[ 2a^2 b \kappa^2 \snrk \cnrk \dnrk\right]^{-1} \frac{\partial \Phi}{\partial \tau} ;\nonumber\\
Y &=& Y_1 \frac{\partial }{\partial \tau} + Y_2 \frac{\partial}{\partial \rho},\nonumber
\eeqa
where we take $(x_1,x_2) = (\tau, \rho)$ in \rf{5.3}.  

For the first solution
\beq
\Phi^{(1)}(\tau,\rho) = 2a^2 \dnrk + \frac{v^4}{4} - a^2(2-\kappa^2)
\label{5.6}
\eeq
in \rf{1.10}, the computation of the corresponding Killing vector field $Y$ is trivial:  By \rf{2.5} and \rf{5.5}  $Y_1 = 2/b$, and of course $Y_2 = 0$.  Since ${\partial g_{11}}/{\partial \tau} = {\partial g_{22}}/{\partial \tau} = 0$ by \rf{4.1}, the Killing equations in \rf{5.3} are satisfied and we see that 
\beq
Y= \frac{2}{b} \frac{\partial}{\partial \tau}\label{5.7}
\eeq
for $(g,\Phi^{(1)})$.  Computations for the other two solutions $\Phi^{(2)}, \Phi^{(3)}$ in \rf{1.10} are more involved.  The result is the following, where again
\beq
A = \frac{1}{16}\left( v^4 - 16v^2a^2 + 8v^2a^2\kappa^2 + 16a^4 \kappa^4\right)
\eeq
in definition \rf{1.11}.

For $\Phi^{(2)}(\tau,\rho)$
\beqa
 Y_1 &=& \frac{  \frac{1}{4} \left( \sinh(\sqrt{A}\tau)\right) \left[ v^2 + (-v^2\kappa^2 - 4a^2 \kappa^4 - 8a^2 \kappa^2)\snSrk + 8a^2 \kappa^4 \sn^4(\rho,\kappa) + 4a^2\kappa^2\right]}{ a^2 b \dnSrk \sqrt{v^2 + (-v^2\kappa^2 - 4a^2\kappa^4)\snSrk+4a^2\kappa^4\sn^4(\rho,\kappa)}}\nonumber\\
 Y_2&=& \frac{ \frac{\sqrt{A}}{4}\left( \cosh(\sqrt{A}\tau)\right) \sqrt{v^2 + (-v^2\kappa^2 - 4a^2\kappa^4)\snSrk+4a^2\kappa^4\sn^4(\rho,\kappa)}}{a^2 b \kappa^2 \snrk\cnrk\dnrk}.\label{5.9}
\eeqa
For $\Phi^{(3)}(\tau,\rho)$ one has quite similar formulas for $Y_1,Y_2$ except (as expected) the roles of the hyperbolic sine and hyperbolic cosine in \rf{5.9} are interchanged:  the factor $\sinh(\sqrt{A}\tau)$ for $Y_1$ in \rf{5.9} is replaced by $\cosh(\sqrt{A}\tau)$, and, similarly, $\cosh(\sqrt{A}\tau)$ for $Y_2$ in \rf{5.9} is replaced by $\sinh(\sqrt{A}\tau)$.

One can also find the following alternative expressions for the Killing vector field components for $\Phi^{(2)}(\tau,\rho)$: 
\beqa
Y_1 &=&  \frac{\left(v^2+4a^2\kappa^2\cnSrk-4a^2\kappa^2\snSrk\right)\sinh(\sqrt{A}\tau)}{4a^2b\dnrk \sqrt{v^2-4a^2\kappa^4 \cnSrk \snSrk \dn^{-2}(\rho,\kappa)}}\label{5.10}\\
 Y_2 &=& \frac{4\sqrt{A}\cosh(\sqrt{A} \tau)\sqrt{v^2-4a^2\kappa^4\cnSrk\snSrk\dn^{-2}(\rho,\kappa)}}{16a^2b\kappa^2\cnrk\snrk}\nonumber
\eeqa
for $A$ in \rf{1.11}.  Corresponding alternative expressions for $\Phi^{(3)}(\tau,\rho)$ are similar to \rf{5.10} except the roles of the hyperbolic sine and hyperbolic cosine are interchanged.  By a direct check one sees that the dilaton fields computed in \rf{1.10} are indeed invariant along the corresponding Killing directions.  That is, they satisfy 
\beq
\frac{\partial \Phi^{(i)}}{\partial \tau}Y_1 + \frac{\partial \Phi^{(i)}}{\partial \rho}Y_2=0
\eeq
for each of $i=1,2,3$ as we indicated in the sentence following equation \rf{5.4} about $Y$ preserving $\Phi$.

\section{Some closing remarks}

For the metric $g$ in \rf{1.6}, whose derivation was discussed in section 2, we have obtained as a main result explicit linearly independent solutions $\Phi^{(j)}$, $j=1,2,3$, in \rf{1.10} of the corresponding system of dilaton field equations in \rf{1.9}.  We have also computed the associated Killing vector fields $Y(g,\Phi^{(j)})$ that leave both $g$ and the $\Phi^{(j)}$ invariant;  see \rf{5.7}, \rf{5.9}, \rf{5.10} and the remarks that follow \rf{5.9} and \rf{5.10}.  The dilaton fields simplify to the expressions given in \rf{4.19} in the special case when the elliptic modulus  $\kappa$ is 1,and $g$ simplifies to the expression given in \rf{4.20}.

For $Q(\rho,\kappa)$ defined in \rf{2.13} it was shown in the short argument following \rf{2.11} that if $Q(\rho,\kappa)=0$ for some $\rho$ then necessarily $v^2/4a^2\kappa^4 \leq 1$:
\beq
|v| \leq 2|a|\kappa^2\label{6.1}
\eeq
in contrast to the standing assumption \rf{1.7}.  To better understand the meaning of this inequality note first by \rf{4.1} that $Q(\rho,\kappa)=0\Rightarrow g_{11}=0$ so $g$ exhibits a horizon singularity at 
\beqa
a^2\kappa^4 \snSrk \cnSrk/\dnSrk &=& \frac{v^2}{4}:\nonumber\\
\snrk\cnrk/\dnrk &=& \pm \frac{v}{2a\kappa^2},\label{6.2}
\eeqa
again by \rf{2.13}.  Keep in mind that $v$ is a velocity parameter -- of a dissipative soliton (also called a \emph{dissipaton}) as in \rf{2.4} for example, especially for $\kappa=1$ as we have remarked at the end of section 2.  The inequality \rf{6.1} is  the statement therefore that for an arbitrary elliptic modulus $\kappa$, with $0<\kappa\leq 1$, the velocity of a black hole dissipaton cannot exceed the limiting value $|v_{max}|\stackrel{def.}{=}2|a|\kappa^2$.  This statement was deduced in \cite{6}, \cite{7}, for example, in the special (but important) case of $\kappa=1$.

In section 3, by a series of explicit transformations of variables, $g$ moreover was transformed to a Jackiw-Teitelboim black hole metric $g_{J-T}$ of the simple form \rf{1.4} -- namely to $g_{J-T}$ given by \rf{3.11}, with accompanying data given by \rf{3.12}, \rf{3.13}.  Here again assumption \rf{1.7} was imposed.  An advantage of the parameterization \rf{3.11} is that, for example, simple formulas exist \cite{10}, \cite{12} for thermodynamic quantities such as the Hawking temperature $T_H$ and black hole entropy $S$.  

We point out, for the record, that the general solutions of all 2d dilaton gravity models are known.  For example, see section 3 of the paper \cite{13} of T.Klosch and T.Stroble.  However (again) we have constructed very explicit elliptic solutions for the specific model of interest - solutions that we feel do not follow directly at all as a corollary of the results of \cite{13}.

In view of an interesting suggestion of the referee, whom we thank (and also for the references \cite{14}, \cite{15}), we add some final remarks that provide a brief review of a connection of the J-T model to cold plasma physics.  This connection is facilitated by way of a \emph{resonant nonlinear Schr\"odinger} (RNLS) equation.

The authors in \cite{14} consider a system of nonlinear equations that describe the dynamics of two-component cold collision-less plasma in the presence of an external magnetic field {\bf B}. For uni-axial plasma propagation, this system is reduced to a system that describes the propagation of nonlinear magneto-acoustic waves in a cold plasma with a transverse magnetic field.  By way of a shallow water approximation of the latter system, a reduction of it to a RNLS equation of the form 
\beq
i\frac{\partial \psi}{\partial t'} + \frac{\partial^2\psi}{\partial {x'}^2} - \frac{1}{2}|\psi|^2\psi = (1+\beta^2)\frac{1}{|\psi|}\frac{\partial^2\psi}{\partial {x'}^2} \psi\label{6.3}
\eeq
is achieved.  Here $x'=\beta x$, $t'=\beta t$ are rescaled space and time variables, and {\bf B} has an expression in terms of a suitable power series expansion in the parameter $\beta^2$.  $\psi$ has the form $\psi=\sqrt{\rho} \ e^{-iS}$ where $S(x',t')$ is a velocity potential and $\rho$ is the mass density of the plasma.  Also note the remarks in \cite{15}.  A key point of interest for us is that for 
\beqa
&r\stackrel{def.}{=}\sqrt{\rho} \ e^{S/\beta}>0,& \stackrel{def.}{=}-\sqrt{\rho} \ e^{S/\beta}<0,\nonumber \\
&B\stackrel{def.}{=}\frac{1}{2\beta^2}=\frac{2}{b^2}, &b\stackrel{def.}{=}2\beta,\label{6.4}
\eeqa
 in the variables ($x',\tau\stackrel{def.}{=}\beta t'$), the reaction diffusion (RD) system \rf{2.1} is satisfied; $r,s$ are denoted by $e^{(+)}, e^{(-)}$ in \cite{14}.  On page 186 of \cite{1} it is shown that, conversely, given solutions $r>0, s<0$ of the RD system \rf{2.1}, one can naturally construct a RNLS solution.  By \rf{2.4}, with $b=2\beta$ by \rf{6.4}, we can take
 \beqa
 r(x',\tau)&=& 2\alpha\beta{\dn }(a(x'-v\tau),\kappa){\rm exp}\left(\left[\frac{v^2}{4}+a^2(2-\kappa^2)\right]\tau-\frac{vx'}{2}\right)\label{6.5}\\
 s(x',\tau)&=& -2\alpha\beta{\dn }(a(x'-v\tau),\kappa){\rm exp}\left(-\left[\frac{v^2}{4}+a^2(2-\kappa^2)\right]\tau+\frac{vx'}{2}\right)\nonumber .
 \eeqa

All of this means that we can apply the prescription \rf{2.3} to construct a metric $g_{\rm plasma}$ of constant Ricci scalar curvature $R=4B\stackrel{def.}{=}\frac{8}{b^2}\stackrel{def.}{=}\frac{2}{\beta^2}$, as we did in \rf{2.7} where the notation $t,\rho$ there is now taken to mean $\tau,a(x'-v\tau)$.  Moreover our results show that $g_{\rm plasma}$ can be transformed to a J-T black hole metric of the form \rf{1.4}.  Thus we can account for a J-T black hole connection in cold plasma physics.  Our results also provide for elliptic solutions of the corresponding dilaton field equations.

\ \\

The authors declare that there is no conflict of interest regarding the publication of this paper.

\end{document}